\documentclass[prl,aps,twocolumn]{revtex4}

\usepackage{psfrag}
\usepackage{graphicx}
\usepackage{dcolumn}
\usepackage{latexsym,amsfonts}
\usepackage{bm}
\usepackage{amssymb}
\begin{document}

\title{Can the ``Darmstadt oscillations'' be treated as two closely
spaced mass--eigenstates of the H--like mother ions ?}

\author{M. Faber$^{a}$\thanks{E-mail: faber@kph.tuwien.ac.at},
A. N. Ivanov${^{\,a,b}}$, P. Kienle$^{b,c}$, E.  L. Kryshen${^d}$,
M. Pitschmann${^a}$, N. I. Troitskaya$^{e}$}
\affiliation{${^a}$Atominstitut der \"Osterreichischen
Universit\"aten, Technische Universit\"at Wien, Wiedner Hauptstrasse
8-10, A-1040 Wien, \"Osterreich} \affiliation{${^b}$Stefan Meyer
Institut f\"ur subatomare Physik \"Osterreichische Akademie der
Wissenschaften, Boltzmanngasse 3, A-1090, Wien,
\"Osterreich}\affiliation{${^c}$Excellence Cluster Universe Technische
Universit\"at M\"unchen, D-85748 Garching, Germany}
\affiliation{${^c}$Petersburg Nuclear Physics Institute, 188300
Gatchina, Orlova roscha 1, Russian Federation}\affiliation{ $^d$
Petersburg Nuclear Physics Institute, 188300 Gatchina, Orlova roscha
1, Russian Federation,}\affiliation{ $^e$State Polytechnic University
of St. Petersburg, Polytechnicheskaya 29, 195251, Russian}
\email{ivanov@kph.tuwien.ac.at}

\date{\today}

\begin{abstract}
  We discuss the proposal that the ``Darmstadt oscillations'' of the
  orbital K--shell electron capture decay $(EC$) rate of the H--like
  heavy ions are caused by quantum beats of two coherently excited,
  closely spaced mass--eigenstates of decaying H--like heavy ions.  We
  show that such a mechanism to explain the time modulation of the
  $EC$--decay rates of the H--like heavy ions leads to the time
  modulation of the $\beta^+$--decay rate of these ions with the same
  period. Such a time--dependence of the $\beta^+$--decay rate of the
  H--like heavy ions contradicts the experimental data of GSI. 
\end{abstract}

\maketitle

Recently Litvinov {\it et al.} \cite{GSI2} have observed that the
K--shell electron capture ($EC$) decay rates of H--like ${^{140}}{\rm
Pr}^{58+}$ and ${^{142}}{\rm Pm}^{60+}$ ions
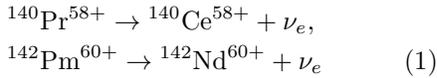
\begin{eqnarray}\label{label1}
&&{^{140}}{\rm Pr}^{58+} \to {^{140}}{\rm Ce}^{58+} + \nu_e,\nonumber\\
&&{^{142}}{\rm Pm}^{60+} \to {^{142}}{\rm Nd}^{60+} + \nu_e
\end{eqnarray}
have an unexpected periodic time modulation of exponential decay
curves. The rates of the number $N^{EC}_d$ of daughter ions
${^{140}}{\rm Ce}^{58+}$ and ${^{142}}{\rm Nd}^{60+}$
\begin{eqnarray}\label{label2}
  \frac{dN^{EC}_d(t)}{dt} = \lambda_{EC}(t)\, N_m(t),
\end{eqnarray}
where $N_m(t)$ is the number of the H--like mother ions ${^{140}}{\rm
Pr}^{58+}$ or ${^{142}}{\rm Pm}^{60+}$\cite{GSI2} and $\lambda^{(\rm
H)}_{EC}(t)$ is the $EC$--decay rate, are periodic functions, caused
by a periodic time--dependence of the $EC$--decay rates
\begin{eqnarray}\label{label3}
  \lambda_{EC}(t) = \lambda_{EC}\,(1 + a_{EC}\, \cos(\omega_{EC}t  +
\phi_{EC}))
\end{eqnarray}
with a period $T_{EC} = 2\pi/\omega_{EC} \simeq 7\,{\rm sec}$, an
amplitude $a_{EC} \simeq 0.20$ and a phase $\phi_{EC}$.

In the articles \cite{Ivanov2,Ivanov3} and the reports
\cite{GSI4,Ivanov5} we have proposed an explanation of the periodic
time--dependence of the $EC$--decay rates as an interference of two
neutrino mass--eigenstates $\nu_1$ and $\nu_2$ with masses $m_1$ and
$m_2$, respectively. The period $T_{EC}$ of the time--dependence has
been related to the difference $\Delta m^2_{21} = m^2_2 - m^2_1$ of
the squared neutrino masses $m_2$ and $m_1$ as follows
\begin{eqnarray}\label{label4}
\omega_{EC} = \frac{2\pi}{T_{EC}} = \frac{\Delta m^2_{21}}{2\gamma M_m},
\end{eqnarray}
where $M_m$ is the mass of the mother ion and $\gamma = 1.43$ is a
Lorentz factor \cite{GSI2}. The analysis of the ``Darmstadt
oscillations'' by means of the mass--differences of neutrino
mass--eigenstates has been carried out also in \cite{Lipkin,Faber} and
\cite{Kleinert}. In a subsequent analysis we also showed that the
$\beta^+$--branches of the decaying H--like heavy ions do not show
time modulation, because of the broad energy spectrum of the neutrinos
in the corresponding three--body decays and proposed a test of such a
behaviour \cite{Ivanov4}.

According to atomic quantum beat experiments \cite{QB1,QB2}, the
explanation of the ``Darmstadt oscillations'', proposed in
\cite{Ivanov2}--\cite{Ivanov5,Faber}, bears similarity with quantum
beats of atomic transitions, when an excited atomic eigenstate decays
into a coherent state of two (or several) lower lying atomic
eigenstates. In the case of the $EC$--decay one deals with a
transition from the initial state $|m\rangle$ to the final state
$|d\,\nu_e\rangle$, where the electron neutrino is a coherent
superposition of two neutrino mass--eigenstates with the energy
difference equal to $\omega_{21} = \Delta m^2_{21}/2 M_m$ related to
$\omega_{EC}$ as $\omega_{EC} = \omega_{21}/\gamma$.

Another mechanism of the ``Darmstadt oscillations'' has been proposed
by Giunti in \cite{Giunti2} and Kienert {\it et al.}
\cite{Lindner}. The authors \cite{Giunti2,Lindner} assume the
existence of two closely spaced mass-eigenstates of the H--like heavy
ion in the initial state of the $EC$--decay and describe the initial
state of the mother ion by the coherent superposition of the wave
functions of two mass-eigenstates
\begin{eqnarray}\label{label5}
|m\rangle = \cos\theta\,|m'\rangle + \sin\theta\,|m''\rangle,
\end{eqnarray}
where $|m'\rangle$ and $|m ''\rangle$ are two mass--eigenstates of the
mother ion with masses $M_{m '}$ and $M_{m ''}$, respectively,
$\theta$ is a mixing angle. By definition of eigenstates the
mass--eigenstates of the H--like heavy ion $|m'\rangle$ and $|m
''\rangle$ should be orthogonal $\langle m'|m''\rangle = 0$.

Unlike our analysis \cite{Ivanov2,Ivanov3,Faber}, the authors
\cite{Giunti2,Lindner} draw an analogy of the ``Darmstadt
oscillations'' with quantum beats of atomic transitions \cite{QB2},
when an atom is excited to a coherent superposition of two closely
spaced upper energy eigenstates and decays into the same lower lying
energy eigenstate. According to \cite{QB2}, the intensity of
radiation, caused by a transition from such a coherent state into the
same lower energy eigenstate, has a periodic time--dependent term with
a period inverse proportional to the energy--difference $\Delta
E_{m'm''}$ between two upper energy eigenstates.

The $EC$--decay rate of the mother ion from the $|m\rangle$ state is
equal to \cite{Ivanov3}
\begin{eqnarray}\label{label6}
\lambda^{(m)}_{EC}(t) = \lambda_{EC}(1 + 2 \sin 2\theta \cos (\Delta
E_{m'm''}t)),
\end{eqnarray}
 where $\lambda_{EC}$ is the $EC$--decay constant
\cite{Ivanov2,Ivanov3,Ivanov1} and $\Delta E_{m'm''}$ is the energy difference
of the mass--eigenstates $|m'\rangle$ and $|m''\rangle$. This shows a
periodic dependence of the $EC$--decay rate with a period inverse
proportional to $\Delta E_{m'm''}$.

However, the H--like heavy ions, subjected to the $EC$--decays, are
unstable also under $\beta^+$--decays \cite{GSI2}: $m \to d + e^+ +
\nu_e$. Following the standard procedure for the calculation of the
$\beta^+$--decay rates \cite{Ivanov3,Ivanov4,Ivanov1} one gets
\begin{eqnarray}\label{label7}
\lambda^{(m)}_{\beta^+}(t) = \lambda_{\beta^+}(1 + 2\sin
2\theta\,\cos(\Delta E_{m'm''}t)),
\end{eqnarray}
where the $\beta^+$--decay constant $\lambda_{\beta^+}$ has been
calculated in \cite{Ivanov1}.  Hence, according to
\cite{Giunti2,Lindner}, the $\beta^+$--decay rates of the H--like
heavy ions should have the same periodic time--dependence as the
$EC$--decay rates. However, this contradicts preliminary experimental
data on the time--dependence of the $\beta^+$--decay rates of the
H--like heavy ${^{142}}{\rm Pm}^{60+}$ ions at GSI, which indicate no
time modulation \cite{GSI3}. This shows that the ``Darmstadt
oscillations'' cannot be treated as quantum beats of two closely
spaced mass--eigenstates of the H--like mother ions.

There is also another reason making a time--dependence of the
$EC$--decay rates of the H--like heavy ions impossible in the
approaches proposed in \cite{Giunti2,Lindner}. For example, the
mass--eigenstates $|m'\rangle$ and $|m''\rangle$ of the mother ions,
injected into the storage sing, would be statistically populated by
the fast projectile fragmentation.  Such a process can also populate
the system of the H--like mother ions with the coherent state
$|\tilde{m}\rangle =-\sin\theta\,|m'\rangle +
\cos\theta\,|m''\rangle$.  Due to statistical equivalence and
principle indistinguishability of these coherent states the
probabilities $P_m$ and $P_{\tilde{m}}$ of the production of the
coherent states $|m\rangle$ and $|\tilde{m}\rangle$, related by $P_m +
P_{\tilde{m}} = 1$, should be equal $P_m = P_{\tilde{m}} =
\frac{1}{2}$.

The decay rate $\lambda^{(\tilde{m})}_{EC}(t)$ and
$\lambda^{(\tilde{m})}_{\beta^+}(t)$ of the $EC$ and $\beta^+$ decays
of the H--like heavy ions from the coherent state $|\tilde{m}\rangle$
are equal to
\begin{eqnarray}\label{label8}
\hspace{-0.3in}\lambda^{(\tilde{m})}_{EC}(t) &=& \lambda_{EC}(1 - 2
\sin 2\theta \cos (\Delta E_{m'm''} t)),\nonumber\\
\hspace{-0.3in}\lambda^{(\tilde{m})}_{\beta^+}(t) &=&
\lambda_{\beta^+}(1 - 2 \sin 2\theta \cos (\Delta E_{m'm''}
t)).
\end{eqnarray}
The total $EC$ and $\beta^+$ decay rates of the H--like heavy ions
from the coherent states $|m\rangle$ and $|\tilde{m}\rangle$ are
defined by
\begin{eqnarray}\label{label9}
\hspace{-0.3in}&&\lambda_{EC}(t) = P_m\lambda^{(m)}_{EC}(t) +
P_{\tilde{m}}\,\lambda^{(\tilde{m})}_{EC}(t) = \lambda_{EC}\nonumber\\
\hspace{-0.3in}&&\times\,(1 + 2 \sin 2\theta\,(P_m -
P_{\tilde{m}}) \cos (\Delta E_{m'm''} t)),\nonumber\\
\hspace{-0.3in}&&\lambda_{\beta^+}(t) = P_m\lambda^{(m)}_{\beta^+}(t) +
P_{\tilde{m}}\,\lambda^{(\tilde{m})}_{\beta^+}(t) = \lambda_{\beta^+}\nonumber\\
\hspace{-0.3in}&&\times\,(1 + 2 \sin 2\theta\,(P_m -
P_{\tilde{m}}) \cos (\Delta E_{m'm''} t)).
\end{eqnarray}
For $P_m = P_{\tilde{m}} = \frac{1}{2}$ one expects no interference
terms in the $EC$ and $\beta^+$ decay rates of the H--like heavy ions.

We would like to accentuate that the analysis of the ``Darmstadt
oscillations'' in terms of the interference of neutrino
mass--eigenstates \cite{Ivanov2,Ivanov3} allows to explain a
time--independence of the $\beta^+$--decay rate of the H--like heavy
ions \cite{Ivanov4}.

We acknowledge fruitful discussions with T. Ericson, F. Bosch and Yu. Litvinov.

\end{document}